\documentclass[12pt]{iopart}
\usepackage{iopams}

\newcommand{\Lie}{\hbox{\pounds}}
\begin{document}

\title{Stationary untrapped boundary conditions in general relativity}

\author{Roh-Suan Tung}
\address{
Center for Astrophysics, Shanghai Normal University,
Shanghai 200234, China}
\ead{tung@shnu.edu.cn}


\begin{abstract}
A class of boundary conditions for canonical general relativity are
proposed and studied at the quasi-local level. It is shown that for
untrapped or marginal surfaces, fixing the {\it area element on the
2-surface} (rather than the induced 2-metric) and the {\it angular
momentum surface density} is enough to have a functionally
differentiable Hamiltonian, thus providing definition of conserved
quantities for the quasi-local regions. If on the boundary the
evolution vector normal to the 2-surface is chosen to be
proportional to the dual expansion vector, we obtain a
generalization of the Hawking energy associated with a generalized
Kodama vector. This vector plays the role for the stationary
untrapped boundary conditions which the stationary Killing vector
plays for stationary black holes. When the dual expansion vector is
null, the boundary conditions reduce to the ones given by the
non-expanding horizons and the null trapping horizons.
\end{abstract}

\section{Introduction}

Traditional description of black holes in terms of event horizons is
inadequate for many physical applications, especially in the cases
of non-stationary spacetimes. Quasi-local notions of trapped and
marginal surfaces have now been found to be more useful for these
cases within the framework of trapping, isolated, and dynamical
horizons \cite{Hayward1994, IH2000, IH2001, IH2002,
Ashtekar-Krishnan2002, Ashtekar-Krishnan2003, Hayward2004,
Booth-Fairhust2004, Gourgoulhon2005, Booth-Fairhust2005,
Hayward2006}. These frameworks enable one to significantly extend
the laws of black hole mechanics to the dynamical regime with the
associated notions of energy, angular momentum and their fluxes, and
have been applied to several problems in mathematical general
relativity, numerical relativity, and quantum gravity
\cite{Ashtekar-Krishnan-Review}.
These progresses on black hole dynamics lead to a question whether
we can generalize the conservation laws for isolated and dynamical
trapping horizons to general untrapped regions so that we can study
the change of energy, angular momentum and their fluxes for
untrapped strong gravitating systems, e.g., before the black hole
horizon was formed.

The question of how to define energy and angular momentum for
untrapped surfaces has been raised for a while in searching for the
``quasi-local energy-momentum and angular momentum''
\cite{Szabados-Review}. The goal has been to find a suitable
definition of total energy-momentum and angular-momentum, surrounded
by a spacelike 2-surface $S$, with $\mathbb{S}^2$ topology, in 4-dimensional
spacetime $M$. The construction is quasi-local in the sense that it
refers only to the geometry of $S$ (intrinsic metric, first
fundamental form), the extrinsic curvatures (second fundamental
forms) and the connection 1-forms on the normal bundle (normal
fundamental forms) for its embedding in $M$.

A systematic way to study conserved quantities is through
the Hamiltonian. For the existence of a functionally differentiable
Hamiltonian for General Relativity, it is necessary to impose
suitable boundary conditions. The allowed boundary conditions for
finite spatial 2-surfaces were studied previously with Dirichlet and
Neumann boundary conditions \cite{Anco-Tung2002a, Anco-Tung2002b}.
This extends the requirement of the functional differentiability of
the Hamiltonian, considered first by Regge and Teitelboim
\cite{Regge-Teitelboim1974} for spatial infinity to the finite
spatial 2-surfaces. Especially interesting boundary conditions are
the Dirichlet boundary conditions which fix the induced metric on
the 2-surfaces.

The conditions were recently relaxed by Szabados who showed that
instead of fixing the full induced metric on the boundary, fixing
the area element is enough. Together with conditions that the lapse
is vanishing on and the shift is tangent to the boundary and is
divergence free, he showed that the Hamiltonian is functionally
differentiable, and that in the large sphere limit the conserved
quantities derived from the Hamiltonian behave as the spatial
components of the total angular momentum \cite{Szabados2006}. This
leads to a question whether or not the condition of vanishing lapse
can be relaxed, and what additional conditions should be imposed so
that a functionally differentiable Hamiltonian can provide a
definition of energy-momentum for the quasi-local region.

In this paper, we propose a class of quasi-local boundary conditions
for canonical general relativity. It is shown that for untrapped or
marginal surfaces, fixing the {\it area element on the 2-surface
$S$} (rather than the induced 2-metric) is enough to have a
functionally differentiable Hamiltonian, thus providing definition of
conserved quantities for the quasi-local regions that allows the
geometry outside to be dynamical and admit gravitational and other
radiation. For gravitating systems including angular momentum, we
further fix the ``{\it angular momentum surface density}'' at
boundary to obtain a generalized definition of quasi-local energy
including angular momentum. These boundary conditions characterize
the equilibrium situation for regions bounded by untrapped or
marginal surfaces.

The evolution vector can be chosen freely. If on the boundary the
evolution vector normal to the 2-surface is chosen to be
proportional to the dual expansion vector, we obtain a
generalization of the Hawking energy associated with a generalized
Kodama vector. This vector plays the role for these stationary
untrapped boundary conditions which the stationary Killing vector
plays for stationary black holes. When the dual expansion vector is
null, the boundary conditions reduce to the ones given by null
trapping horizons and non-expanding horizons.

We begin with the geometry of an untrapped 2-surface $S$ embedded in
a 4-dimensional spacetime $M$. Introduce a set of orthonormal vectors
$e_0, e_1, e_2, e_3$ adapted to the 2-surface $S$, with $e_0$ and
$e_1$ being the set of timelike and spacelike unit normals to $S$ and
$e_A=[e_2, e_3]$ being tangent to $S$. The extrinsic curvatures of $S$
with respect to $e_0$ and $e_1$ directions are given by
$k(e_0)_{AB}=g(e_B ,\nabla_A e_0)=-\Gamma_{0BA}$ and $k(e_1)_{AB}=
g(e_B, \nabla_A e_1)=-\Gamma_{1BA}$. The connection 1-forms in the
normal bundle are given by $\varpi_{A}=g(e_1, \nabla_A
e_0)=-\Gamma_{01A}$. Here $\Gamma_{IJK}= -g(e_J, \nabla_K e_I)$ are
Ricci rotation coefficients. Under a boost transformation of $e_0$
and $e_1$,
\begin{eqnarray} \label{boost-e}
{e'}_0 &=& e_0 \cosh u + e_1 \sinh u ,\nonumber\\
{e'}_1 &=& e_0 \sinh u + e_1 \cosh u ,
\end{eqnarray}
they transform as
\begin{eqnarray} \label{boost-kw}
k({e'}_0)_{AB} &=& k(e_0)_{AB} \cosh u + k(e_1)_{AB} \sinh u , \nonumber\\
k({e'}_1)_{AB} &=& k(e_0)_{AB} \sinh u + k(e_1)_{AB} \cosh u ,
\nonumber\\
{\varpi'}_A &=& \varpi_A - \nabla_A u .
\end{eqnarray}

The expansion vector $H$, and the dual expansion vector $H_\perp$
are defined with the trace of the extrinsic curvatures $k(e_0)$ and
$k(e_1)$,
\begin{eqnarray} \label{expansionvector}
H &=& k(e_1) e_1 - k(e_0) e_0  ,\\
H_\perp &=& k(e_1) e_0 - k(e_0) e_1  , \label{dualexpansionvector}
\end{eqnarray}
where $H$ and $H_\perp$ are also known as the mean curvature vector
and the dual mean curvature vector respectively. These vectors are
independent of choice of normal frames for the 2-surface. They are
invariant under the boost transformation (\ref{boost-e}). Thus, they
depend only on the 2-surface $S$ and constitute a set of natural
normal vectors for $S$ \cite{Szabados1994}.

A 2-surface $S$ is {\it trapped} if $k(e_1)^2>k(e_0)^2$, {\it
untrapped} if $k(e_1)^2<k(e_0)^2$, and {\it marginal} if
$k(e_1)^2=k(e_0)^2$, everywhere on $S$. Untrapped surfaces are also
called {\it mean convex} surfaces. The dual expansion vector
$H_\perp$ is always timelike for untrapped surfaces, null for
marginal surfaces \cite{Szabados1994, Anco-Tung2002a,
Anco-Tung2002b}. Note that, on $S$, the trace of the extrinsic curvature is
zero along the direction of the dual expansion vector, i.e.
$k(H_\perp)\vert_S=0$, thus we have
\begin{equation} \label{expansionzero}
\Lie_{H_\perp} \varrho \vert_S=0 ,
\end{equation}
where $\varrho$ is the area element of $S$.

\section{Conserved quantities derived from a hamiltonian}

For a general diffeomorphism-invariant field theory in four
dimensions with a Lagrangian 4-form ${\cal L}(\phi)$, where $\phi$
denotes an arbitrary collection of dynamical fields, the field
equations, ${\cal E}=0$, are obtained by computing the first
variation of the Lagrangian,
\begin{equation}
\delta {\cal L}= d \Theta (\phi , \delta\phi ) + {\cal E} \delta
\phi ,
\end{equation}
where $\Theta (\phi,\delta\phi)$ is the symplectic potential 3-form.
For any diffeomorphism generated by a smooth vector field $\xi$,
\begin{equation}
d i_\xi {\cal L} =\Lie_\xi {\cal L} = d\Theta(\phi, \Lie_\xi \phi) +
{\cal E} \Lie_\xi \phi ,
\end{equation}
where ${\rm\pounds}_\xi$ denotes the Lie derivative and $i_\xi$ is
the inner product, one can define a conserved Noether current 3-form
$J(\xi)$ by
\begin{equation}
J(\xi)=\Theta(\phi,{\rm\pounds}_\xi\phi)-i_\xi {\cal L}(\phi) ,
\end{equation}
such that the Noether current $J(\xi)$ is closed ($dJ(\xi)=-{\cal E}
\Lie_\xi \phi \simeq 0$) when the field equations are satisfied.
Locally there exist a 2-form $Q(\xi)$ (called the Noether charge)
such that $J(\xi)=d Q(\xi)$. The variation of the Noether current
3-form is given by
\begin{equation}
\delta J(\xi)= \omega(\phi, \delta\phi, {\rm\pounds}_\xi\phi) + d
( i_\xi \Theta (\phi,\delta\phi ) ) ,
\end{equation}
where $\omega$ is the symplectic current 3-form defined by
\begin{equation}
\omega(\phi,
\delta_1\phi,\delta_2\phi)=\delta_1\Theta(\phi,\delta_2\phi)-
\delta_2\Theta(\phi,\delta_1\phi) .
\end{equation}
Its integral over a 3-surface $\Sigma$ defines the presymplectic
form $\Omega$. If the presymplectic form is a total variation
\begin{eqnarray}
\Omega(\phi, \delta\phi, \Lie_\xi\phi) &\equiv&
\int_\Sigma\omega(\phi,\delta\phi,\Lie_\xi\phi)  =\delta
\mathbb{H}(\xi) ,
\end{eqnarray}
for some function $\mathbb{H}(\xi)$ on the field space, then
$\mathbb{H}(\xi)$ is conserved along $\xi$, i.e. we have $\Lie_\xi
\mathbb{H}(\xi)=0$. The function $\mathbb{H}(\xi)$ is called the
Hamiltonian conjugate to $\xi$
\cite{Lee-Wald1990,Nester1991,Wald1993,Wald2000,Chen-Nester-Tung2005}.
Note that on shell, the presymplectic form is given by
\begin{equation} \label{symplectic2}
\int_\Sigma\omega(\phi,\delta\phi,\Lie_\xi\phi) = \int_\Sigma
\delta J(\xi)-d(i_\xi\Theta) =\oint_S \delta Q(\xi) -
i_\xi\Theta .
\end{equation}

For General Relativity we begin with the Hilbert action,
\begin{equation}
 S=\int {\cal L}=\int R^{ab}\wedge
 \ast(\vartheta_a\wedge\vartheta_b) ,
 \end{equation}
where $R^{ab}=d\Gamma^{ab}+\Gamma^a{}_c \wedge \Gamma^{cb}$ is the
curvature 2-form constructed by the connection 1-form $\Gamma^{ab}$,
 $\ast(\vartheta^a\wedge\vartheta^b)=\frac{1}{2}
 \epsilon^{ab}{}_{cd}\vartheta^c\wedge\vartheta^d$, and $g=
 \eta_{ab}\, \vartheta^a \otimes_s \vartheta^b$ is the metric,
 where $\eta_{ab}={\rm diag}(-1,1,1,1)$ and
 $\vartheta^a$ is the orthonormal
 frame 1-form field. A variation of the Lagrangian gives,
\begin{eqnarray}
 \delta {\cal L}
 &=& R^{ab}\wedge \epsilon_{abcd} \vartheta^c
 \wedge \delta \vartheta^d
 + \delta \Gamma^{ab} \wedge D \ast(\vartheta_a\wedge\vartheta_b)
 \nonumber\\
 && + d \left(\delta \Gamma^{ab} \wedge \ast(\vartheta_a\wedge\vartheta_b) \right) ,
 \end{eqnarray}
which identifies the symplectic potential $\Theta(\phi,\delta\phi)
 =\delta \Gamma^{ab}\wedge \ast(\vartheta_a\wedge\vartheta_b)$.
The Noether current 3-form is given by
\begin{eqnarray}
J(\xi) &=& d Q(\xi) = d [ (i_\xi \Gamma^{ab}) \,
\ast(\vartheta_a\wedge\vartheta_b) ]  ,
\end{eqnarray}
 where $Q(\xi)$ is the Noether charge 2-form and we assume the
 field equations are satisfied. The presymplectic form is
 given by
\begin{equation}
\int_\Sigma \omega =\oint_S \delta Q(\xi)-i_\xi\Theta=\oint_S
C_1(\xi)+\oint_S C_2(\xi) ,
\end{equation}
 where, for convenience, we define
\begin{equation}
 C_1(\xi)= \frac{1}{2} i_\xi \Gamma^{ab}
\delta(\epsilon_{abcd}\vartheta^c\wedge\vartheta^d ) ,
\end{equation}
 and
\begin{equation}
 C_2(\xi)=
i_\xi\vartheta^c\wedge\delta\Gamma^{ab}\wedge
\epsilon_{abcd}\vartheta^d .
\end{equation}
Let us first expand $C_1$ to the normal and
tangent components,
\begin{eqnarray}
\oint_S C_1(\xi) &=&\oint_S i_\xi \Gamma^{01} \delta
(\epsilon_{01AB} \vartheta^A\wedge\vartheta^B) +
\oint_S i_\xi \Gamma^{AB} \delta (\epsilon_{AB01}
\vartheta^0\wedge\vartheta^1) \nonumber\\
&&+2 \oint_S i_\xi \Gamma^{0A} \delta (\epsilon_{0A1B}
\vartheta^1\wedge\vartheta^B)
+ 2 \oint_S i_\xi \Gamma^{1A} \delta (\epsilon_{1A0B}
\vartheta^0\wedge\vartheta^B)  .
\end{eqnarray}
Let $P_S$ be the projection onto $S$, we have
$P_S\vartheta^0=P_S\vartheta^1=0$. Moreover, assuming
$\delta(P_S)=0$, we have
\begin{equation}
P_S\delta\vartheta^0=\delta(P_S\vartheta^0)-(\delta
P_S)\vartheta^0=0 ,
\end{equation}
and likewise $P_S\delta\vartheta^1=\delta(P_S\vartheta^1)-(\delta
P_S)\vartheta^1=0$. Thus the term with $C_1$ reduces to
\begin{equation} \label{e21}
\oint_S C_1(\xi)=2 \oint_S i_\xi\Gamma^{01} \delta\varrho ,
\end{equation}
where $\varrho=\frac{1}{2}(\epsilon_{01AB}
\vartheta^A\wedge\vartheta^B)$ is the area element on $S$.

Similarly for $C_2$, by a projection onto the 2-surface $S$, it is
straightforward to show that the following identity holds on $S$ for
any vector field $V=V^a e_a$:
\begin{equation} \label{112identity}
V^c \, \delta\Gamma^{ab} \wedge \epsilon_{abcd} \vartheta^d
\vert_S
=- 2\, \varrho \, (V^0 \delta k(e_1) + V^1 \delta k(e_0)
- V^A \delta \varpi_A )\vert_S .
\end{equation}
Using this identity, we obtain
\begin{equation} \label{e23}
\oint_S C_2(\xi)
  = -\oint_S 2 \varrho \, \left(i_\xi \vartheta^0 \delta k(e_1) + i_\xi \vartheta^1
\delta k(e_0)
 -i_\xi \vartheta^A \delta\varpi{}_A \right)  .
\end{equation}
By (\ref{e21}) and (\ref{e23}), the full symplectic form is,
\begin{eqnarray} \label{18}
\int_\Sigma\omega &=&\oint_S C_1(\xi)+C_2(\xi) \nonumber\\
&=& \oint_S 2 \, i_\xi\Gamma^{01} \delta\varrho  -
\oint_S 2 \varrho \left(i_\xi \vartheta^0 \delta k(e_1) +i_\xi
\vartheta^1 \delta k(e_0)
 -i_\xi \vartheta^A \delta\varpi{}_A \right)  .
 \end{eqnarray}
This is the key equation for our discussion.

\section{Energy}

We assume that the timelike (or null) vector $\xi$ is fixed on $S$,
\begin{equation}
\delta \xi \vert_S=0 , \qquad \hbox{(Boundary Condition I),}
\end{equation}
moreover, we assume
  \begin{equation}
\delta \varrho \vert_S=0 , \qquad \hbox{(Boundary Condition II),}
\label{bc2}
 \end{equation}
i.e. the area element of the 2-surface $\varrho$ is fixed.

We first consider a special case by assuming that, on $S$, $i_\xi
\vartheta^A \vert_S=0$ (this condition will be relaxed in the next
section). Using the boundary conditions (I, II),
\begin{equation}
  \oint_S C_2(\xi)=  - \delta \oint_S \varrho \, i_\xi
  \left( \vartheta^0 k(e_1)+ \vartheta^1 k(e_0) \right)  ,
\end{equation}
is a total variation, and
\begin{equation}
\oint_S C_1(\xi)=2 \oint_S i_\xi\Gamma^{01} \delta\varrho=0 .
\end{equation}
Thus with boundary conditions (I) and (II), the full symplectic
form,
\begin{equation}
 \int_\Sigma\omega
 = - \delta \oint_S \varrho \, i_\xi
  \left( \vartheta^0 k(e_1)+ \vartheta^1 k(e_0) \right) =\delta \mathbb{E}_H(\xi),
 \end{equation}
is a total variation. The Hamiltonian $\mathbb{E}_H(\xi)$ associated
with the vector $\xi$ is given by
\begin{equation} \label{e30}
 \mathbb{E}_H(\xi)
 =   \oint_S  \, \left(f(\varrho)-  i_\xi
  \left( \vartheta^0 k(e_1)+ \vartheta^1 k(e_0) \right) \right) \varrho .
\end{equation}
where $f(\varrho)$ is a function of the area element. The evolution
vector $\xi$ can be chosen freely. By choosing $\xi$ to be timelike
or spacelike vector, the Hamiltonian gives energy or momentum
respectively. Here we assume $\xi$ is timelike.

Note that because of equation (\ref{expansionzero}), the boundary
conditions {(I)} and {(II)} are both satisfied if we replace the
variation with the Lie derivative with respect to $\xi$
($\delta=\Lie_\xi$), and assume that on $S$, the evolution vector
$\xi$ is given by
\begin{equation} \label{kodama}
\xi\vert_S=h(\varrho) H_\perp ,
\end{equation}
where $H_\perp$ is the dual expansion vector and $h(\varrho)$ is a
function of the area element on the (untrapped or marginal)
2-surface $S$.

The boundary condition II implies that the area is conserved along
the dual expansion vector. Since the area of each cross section does
not change along the dual mean curvature vector direction, we called
it the {\em stationary untrapped boundary conditions}.

Note that because of equation (\ref{kodama}),
$\xi\vert_S=h(\varrho)H_\perp$, so by equation
(\ref{dualexpansionvector}), we have, on $S$, $i_\xi
\vartheta^0=h(\varrho) k(e_1)$, $i_\xi \vartheta^1=-h(\varrho)
k(e_0)$, and $i_\xi \vartheta^A=0$. Thus the Hamiltonian
$\mathbb{E}_H(\xi)$ associated with the vector $\xi$ is given by
\begin{equation}
 \mathbb{E}_H(\xi)
 =   \oint_S  \, \left(f(\varrho)-h(\varrho) H^2 \right) \varrho .
\end{equation}

The free functions of the area element, $f(\varrho)$ and
$h(\varrho)$, can be chosen such that the expression gives ADM mass
at spatial infinity and irreducible mass at marginal surfaces $H=0$.
This can be done by letting $f(\varrho)$ to be $1/(8\pi \mathbb{R})$
and let $h(\varrho)$ to be $\mathbb{R}/(32\pi)$, where $\mathbb{R}$
is the area radius given by
\begin{equation}
\mathbb{R}=\sqrt{\frac{1}{4\pi}\oint_S \varrho} .
\end{equation}
This leads to the energy expression,
\begin{equation}
 \mathbb{E}_H(\xi)
 = \frac{\mathbb{R}}{2} \left( 1 - \frac{1}{16\pi} \oint_S  \, \,
 H^2 \varrho \right) ,
\end{equation}
which is precisely the Hawking energy \cite{Hawking1968} .

\section{Angular momentum}

The equation (\ref{18}) can also be used to define angular momentum.
Let $\xi\vert_S=(8 \pi)^{-1} \psi $ be a vector tangent to $S$
satisfying $i_\psi \vartheta^0 \vert_S=0$ and $i_\psi \vartheta^1
\vert_S=0$, and
\begin{equation}
\delta\psi \vert_S=0, \qquad \hbox{(Boundary Condition I$'$),}
\end{equation}
and
\begin{equation}
\varpi_A \delta \vartheta^A \vert_S=0, \qquad \hbox{(Boundary
Condition A),}
\end{equation}
then
\begin{equation}
\int_\Sigma \omega = \frac{1}{8\pi} \delta \oint_S \psi^A\varpi_A
\varrho = \delta \mathbb{J}(\psi) ,
\end{equation}
is a total variation, where $\psi^A=i_\psi \vartheta^A$. The
Hamiltonian associated with $\psi$ is the angular momentum given by
\begin{equation}
\mathbb{J}(\psi)=\frac{1}{8\pi} \oint_S j \, \varrho=\frac{1}{8\pi}
\oint_S \psi^A \varpi_A \,\varrho ,
\end{equation}
with the ``angular momentum surface density'' given by $j= \psi^A
\varpi_A$.

Because of the gauge freedom in choosing the normal fundamental forms
$\varpi_A$ (equation (\ref{boost-kw})), the definition of angular
momentum is not unique. A further condition which makes it unique is
\begin{equation}
\delta_\psi \varrho \vert_S =\Lie_\psi \varrho \vert_S=0, \qquad
\hbox{(Boundary Condition II$'$)}.
\end{equation}
Here $\psi$ generates a symmetry of the area form rather than the
whole metric. This implies that $\psi$ has vanishing transverse
divergence $\nabla_A\psi^A=0$. Under the gauge freedom for the
normal fundamental forms $\varpi_A \mapsto \varpi_A - \nabla_A u$,
the angular momentum formula
\begin{equation}
\oint_S \psi^A\varpi_A \mapsto \oint_S\psi^A (\varpi_A - \nabla_A
u)
= \oint_S \psi^A \varpi_A - \nabla_A(\psi^A u)
=\oint_S \psi^A \varpi_A ,
\end{equation}
is invariant. Thus the angular momentum formula is uniquely defined
on $S$ (up to the choice of $\psi^A$). This condition is also given
in \cite{Ashtekar-Krishnan2003, Booth-Fairhust2005, Gourgoulhon2005,
Hayward2006} for dynamical trapping horizons and was also used by
Szabados \cite{Szabados2006} as a Hamiltonian boundary condition for
quasi-local angular momentum.

We can now extend the stationary untrapped boundary conditions to
include angular momentum.  Similar to the discussion in the previous
section, we require that
 $\delta \xi \vert_S = 0$  {(Boundary Condition I)} and
$\delta \varrho \vert_S=0$ {(Boundary Condition II)} are satisfied.
In addition, we further require that Boundary Condition A is
satisfied. This implies that the angular momentum surface density
$j$ is fixed on $S$, i.e.
 \begin{equation}
\delta j \vert_S=0 , \qquad \hbox{(Boundary Condition III).}
\label{bc4}
 \end{equation}

The full symplectic form is then given by applying the boundary
conditions to the equation (\ref{18}),
\begin{equation}
\int_\Sigma \omega(\phi,\delta\phi, \Lie_\xi\phi)=\delta
\mathbb{E}(\xi),
\end{equation}
where the Hamiltonian associated with $\xi$ is now given by
\begin{equation} \label{EnJ}
 \mathbb{E}(\xi)
 =   \oint_S  \, \left(f(\varrho, j) -  i_\xi
  \left( \vartheta^0 k(e_1)+ \vartheta^1 k(e_0) \right)
  \right) \varrho .
\end{equation}

A natural choice of the evolution vector on $S$ is
\begin{equation} \label{fullvector}
\xi\vert_S=h(\varrho, j) H_\perp - \Omega(\varrho, j)  \psi ,
\end{equation}
which is assumed to be timelike or null. The free functions
 $h(\varrho, j)$ and $\Omega(\varrho, j)$ (angular speed) are now
functions of the area element $\varrho$ and
the angular momentum surface density $j$.

 Now we replace the variation $\delta$ by the Lie derivative
 $\Lie_\xi$, the boundary condition (I) is automatically satisfied
 ($\Lie_\xi \xi=0$). By equation (\ref{fullvector}),
 the boundary condition (II) is
satisfied with the following condition on $S$:
\begin{equation} \label{condition1}
\Lie_\psi \varrho \vert_S=0 .
\end{equation}
The boundary condition (III) is satisfied if the vector field $\psi$
is transported by $H_\perp$,
\begin{equation} \label{condition2}
\Lie_\psi {H_\perp} \vert_S=-\Lie_{H_\perp} \psi \vert_S = 0 .
\end{equation}
The conditions (\ref{condition1}) and (\ref{condition2}) are
consistent with the ones in \cite{Ashtekar-Krishnan2003,
Gourgoulhon2005, Booth-Fairhust2005, Hayward2006} for dynamical
trapping horizons. These conditions then imply that
\begin{equation}
\Lie_\xi \xi \vert_S=0, \qquad \Lie_\xi \varrho \vert_S=0, \qquad
\Lie_\xi j \vert_S= 0 ,
\end{equation}
the stationary untrapped boundary conditions (I,II,III) are
satisfied.

By equation (\ref{EnJ}), the Hamiltonian associated with the
evolution vector (\ref{fullvector}) is then given by
\begin{equation}
 \mathbb{E}(\xi)
 =   \oint_S  \, \left(f(\varrho, j)-h(\varrho, j) H^2) \right)
 \varrho .
\end{equation}
By requiring that the energy expression gives the standard value for Kerr
black holes at marginal surface $H=0$,
\begin{equation}
\mathbb{E}_{horizon}=\frac{\sqrt{\mathbb{R}^4+4 \mathbb{J}^2}}{2
\mathbb{R}} ,
\end{equation}
we obtain
\begin{equation}
f(\varrho,j)=\frac{\sqrt{\mathbb{R}^4+4 \mathbb{J}^2}}{{8\pi
\mathbb{R}^3}} .
\end{equation}
The other free function $h(\varrho, j)$ can be chosen such that the
energy is proportional to the Hawking energy and gives ADM mass at
spatial infinity, this implies that
\begin{equation}
h(\varrho,j)=\frac{\sqrt{\mathbb{R}^4+4 \mathbb{J}^2}}{32 \pi
\mathbb{R}} .
\end{equation}
Then the Hamiltonian associated with $\xi$,
\begin{equation}
 \mathbb{E}(\xi)
 =\frac{\sqrt{\mathbb{R}^4+4 \mathbb{J}^2}}{2 \mathbb{R}}
 \left( 1 - \frac{1}{16\pi} \oint_S  \, \, H^2
 \varrho \right) ,
 \end{equation}
provides a suitable choice for the energy expression.

Note that the vector on $S$,
\begin{equation} \label{generalized-kodama}
\xi \vert_S= \frac{\sqrt{\mathbb{R}^4+4 \mathbb{J}^2}}{32 \pi
\mathbb{R}} H_\perp - \Omega \psi ,
\end{equation}
reduces to the Kodama vector
\begin{equation}
\xi_{Kodama}= \frac{\mathbb{R}}{32 \pi} H_\perp ,
\end{equation}
in spherically symmetric spacetimes \cite{Kodama1980}. Thus equation
(\ref{generalized-kodama}) is a generalized Kodama vector for
non-spherically symmetric spacetimes.

\section{Discussion}

In summary, for quasi-local regions bounded
by an untrapped or marginal 2-surface,
a functional differentiable Hamiltonian can be defined with
the boundary conditions which fix
the evolution vector (equation (25)),
the area element (equation (26))
and the angular momentum surface density (equation (41)) on the 2-surface.
As a consequence, a preferred expression of quasi-local energy for
these ``stationary untrapped boundary conditions" is given,
which allows the geometry outside to be dynamical and
to admit gravitational and other radiation.
The quasi-local energy expression (equation (52)) generalizes the Hawking energy
to include angular momentum. On the boundary, the evolution vector
associated with this expression is a generalization of the Kodama vector
(equation (53)). These results generalize the conserved quantities for Isolated
Horizons so as to provide covariant conserved quantities
for general untrapped regions.

In equation (31), the evolution vector on the boundary is chosen
to be proportional to the dual expansion vector. Alternatively,
we can use the unit dual expansion vector.
If $H$ is not null ($|H|\neq0$), there is a set of uniquely
determined unit normal vectors for the 2-surface, given by
\begin{equation} \label{preferrednormals}
\hat{e}_{0} = \frac{H_\perp}{|H|} \qquad \hat{e}_{1} = \frac{H}{|H|}
,
\end{equation}
where  $|H|=\sqrt{H\cdot H}=\sqrt{k(e_1)^2-k(e_0)^2}$.
If we choose the evolution vector $\xi$
such that $\xi \vert_S=\hat{e}_0$, then
by equation (\ref{e30}), this leads to the energy expression
\begin{equation}
 \mathbb{E}(\xi)
 =   \oint_S  \, \left(f(\varrho)- \sqrt{k(e_1)^2-k(e_0)^2}) \right)
 \varrho .
\end{equation}
Here we have only one free function $f(\varrho)$. A natural requirement is
that the expression should give ADM energy at spatial infinity. 
Assuming that we
can embed the 2-surface isometrically into Minkowski spacetime,
let  $k_0(e_1)$ be the trace of extrinsic curvature with respect to $e_1$,
for the 2-surface in Minkowski spacetime, then the choice
$f(\varrho)=k_0(e_1)$ gives the  Kijowski-Liu-Yau quasi-local energy
\cite{Kijowski1997, Liu-Yau2003}. In the special  $k(e_0)=0$ slice,
the Kijowski-Liu-Yau energy equals the Brown-York energy
\cite{Brown-York1993, Brown-Lau-York2002}. There are
nice positive energy theorems for these expressions
\cite{Liu-Yau2003,Shi-Tam2002}.  However,
$\hat{e}_{0}$ fails to be defined in the null case. Moreover,
the requirement of isometric embedding assumes that the full
induced 2-metric is fixed (rather than just the area element).
Thus it seems that this choice is not suitable for the cases
involving dynamical black holes.

The stationary untrapped boundary conditions can be compared with
the Isolated Horizon boundary conditions. The untrapped surface $S$
together with the associated evolution vector on the boundary
$\xi\vert_S$ constitute a timelike hypersurface.
In the limit when the dual expansion vector
$H_\perp$ is null, $S$ reduces to a marginal surface, the
hypersurface reduces to a non-expanding horizon (null 3-surfaces in
spacetime on which the expansion with respect to any null normal is
zero) \cite{Ashtekar-Krishnan-Review}. This suggests a
generalization of the non-expanding horizons to ``{\em stationary
untrapped hypersurfaces}'' (with the boundary conditions I, II, III)
describing the equilibrium states of untrapped surfaces with
conservation of the area (with area radius $\mathbb{R}$),  the
energy $\mathbb{E}$, and the angular momentum $\mathbb{J}$.

\bigskip

\ack
The author thanks Sean A. Hayward,
James M. Nester and Laszlo B. Szabados for helpful discussions. This
work was supported by the National Natural Science Foundation of
China under grant numbers 10375081, 10375087 and 10771140, by
Shanghai Pujiang Talent Program, by Shanghai Education Development
Foundation Shuguang Award and by NCTS Taiwan.


\end{document}